\documentclass{llncs}
\usepackage{graphicx}
\usepackage{subfigure}
\usepackage{enumitem}

\usepackage{subfigure}
\usepackage{color}
\usepackage{algorithm}
\usepackage{algorithmicx}

\newcommand{\qmarks}[1]{``#1''}

\begin{document}
\title{Adventures in Abstraction: Reachability in Hierarchical Drawings}
\author{Panagiotis Lionakis\inst{1,2}  \and
Giacomo Ortali\inst{3}\and Ioannis G. Tollis\inst{1,2}}
\authorrunning{P. Lionakis et al.}
\institute{Computer Science Department, University of Crete, GREECE \email{{\{lionakis,tollis\}@csd.uoc.gr}}\\ \and
Tom Sawyer Software, Inc. Berkeley, CA 94707 U.S.A
\\ \and
University of Perugia
\email{giacomo.ortali@gmail.com}\\}

\maketitle              % typeset the header of the contribution
\begin{abstract}
We present algorithms and experiments for the visualization of directed graphs that focus on
displaying their reachability information. Our algorithms are based on the concepts of the path and channel decomposition as proposed in the framework presented in~\cite{ortali2018algorithms} and focus on showing the existence of paths clearly.  In this paper we customize these concepts and present experimental results that clearly show the interplay between bends, crossings and clarity.   Additionally, our algorithms have direct applications to the important problem of showing and storing transitivity information of very large graphs and databases. Only a subset of the edges is drawn, thus reducing the visual complexity of the resulting drawing, and the memory requirements for storing the transitivity information.
Our algorithms require almost linear time, $O(kn+m)$,  where $k$ is the number of paths/channels, $n$ and $m$ is the number of vertices and edges, respectively.  They produce progressively more abstract drawings of the input graph.  No dummy vertices are introduced and the vertices of each path/channel are \emph{vertically aligned}.

\end{abstract}
\section{Introduction}
The visualization of directed (sometimes acyclic) graphs has many applications in several areas of science and business. Such graphs often represent hierarchical relationships between objects in a structure (the graph).  In several applications, such as graph databases and big data, the graphs are very large and the usual visualization techniques are not applicable.  In their seminal paper of 1981, Sugiyama, Tagawa, and Toda~\cite{sugiyama1981methods} proposed a four-phase framework for producing hierarchical drawings of directed graphs. This framework is known in the literature as the \qmarks{Sugiyama} framework, or algorithm. Most problems involved in the optimization of various phases of the Sugiyama framework are NP-hard.  In~\cite{ortali2018algorithms} a new framework is introduced  to visualize directed graphs and their hierarchies which departs from the classical four-phase framework of Sugiyama and computes readable hierarchical visualizations by \qmarks{hiding} (\emph{abstracting}) some selected edges while maintaining the complete reachability information of a graph.  
\par
In this paper we present several algorithms that follow that framework.  Our algorithms reduce the visual complexity of the resulting drawings by (a) drawing the vertices of the graph in some vertical lines, and (b) by progressively \emph{abstracting} some transitive edges thus showing only a subset of the edge set in the output drawing.  The process of progressively abstracting the edges gives different visualization results, but they all have the same transitive closure as the input graph.  Notice that this type of abstraction has additional applications in storing the transitive closure of huge graphs, which is a significant problem in the area of graph databases and big data~\cite{Jagadish:1990:CTM:99935.99944,DBLP:conf/sigmod/JinRDY12,DBLP:conf/sigmod/SchaikM11,DBLP:conf/edbt/VelosoCJZ14,DBLP:journals/vldb/YildirimCZ12}.  We also present experimental results that show a very interesting interplay between bends, crossings, clarity of the drawings, and the abstraction of edges.

A \emph{path} and a \emph{channel} are both ordered sets of vertices. In a path every vertex is connected by a direct edge to its successor, while in a channel any vertex is connected to it by a directed path (which may be a single edge). The concept of channel can be seen as a generalization of the concept of path. In the literature the channels are also called \emph{chains}~\cite{Jagadish:1990:CTM:99935.99944}.

\begin{figure}[!htp]
\centering
\subfigure[]{
    \includegraphics[width=0.3\textwidth,height=12cm]{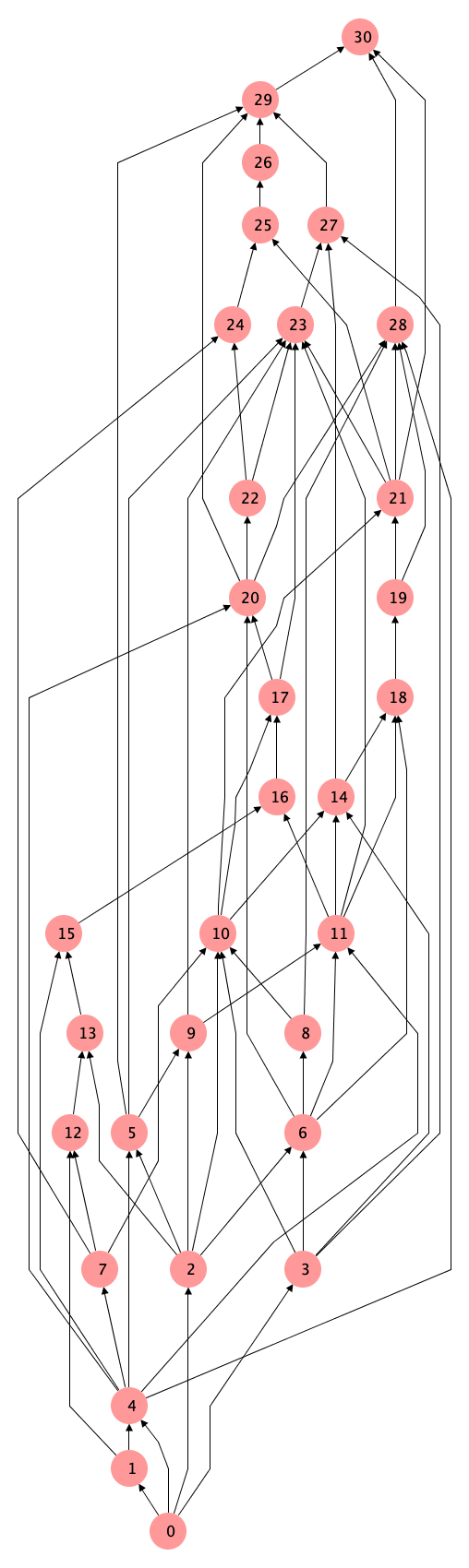}
}
\subfigure[]{
    \includegraphics[width=0.3\textwidth,height=12cm]{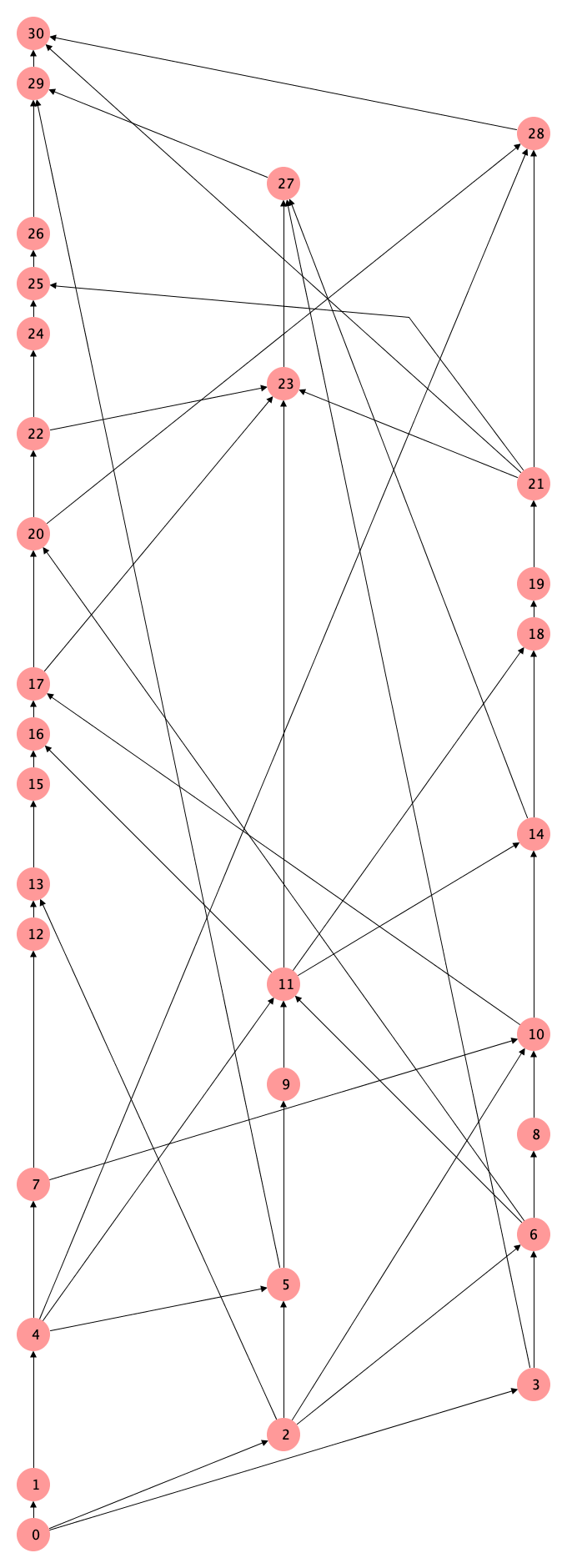}
    \label{fig:1}
}
\subfigure[]{
    \includegraphics[width=0.3\textwidth,height=12cm]{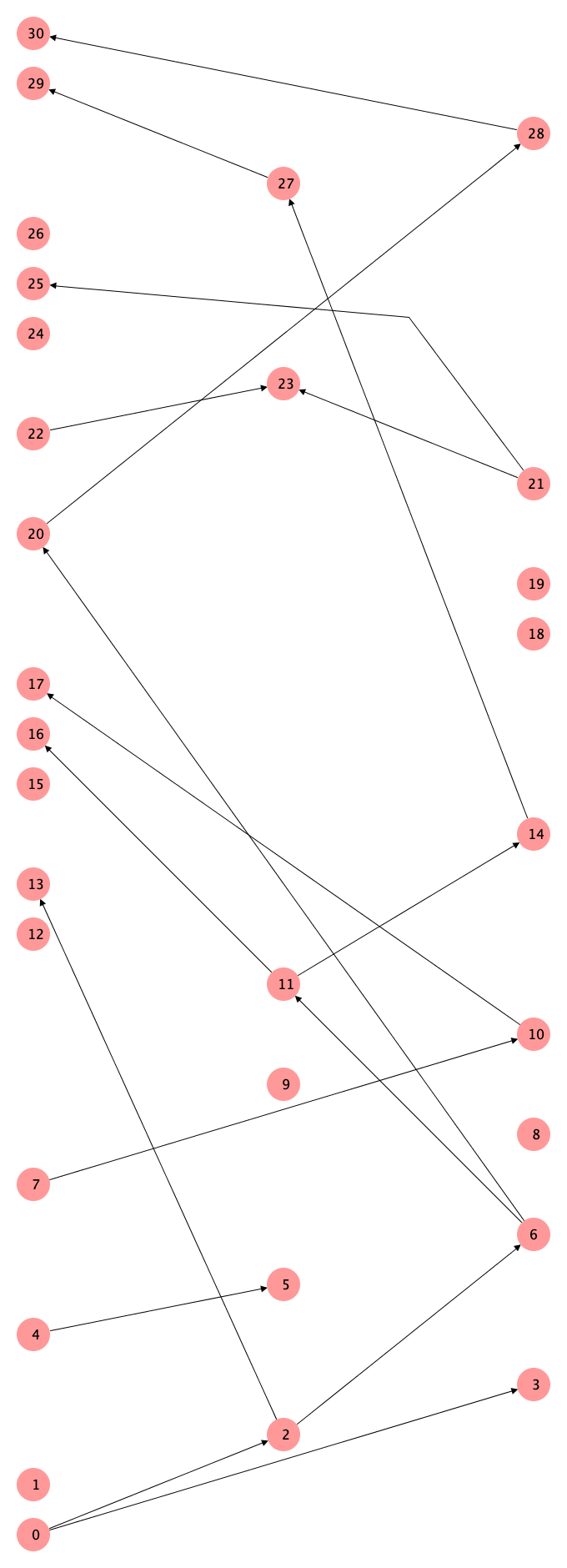}
}
\caption[]{In (a) we show the drawing of a graph $G$ as computed by Tom Sawyer Perspectives which follows the Sugiyama framework. In (b) we show the drawing $\Gamma$ based on $G$ computed by Algorithm PB-Draw. In (c) we show an abstracted hierarchical drawing computed by our final variant.}
\label{teaser}
\end{figure}

Figure~\ref{teaser} shows an example of three different hierarchical drawings: part~(a) shows the drawing of a directed graph $G$ computed by Tom Sawyer Perspectives~\cite{Tom} (a tool of Tom Sawyer Software) that follows the Sugiyama framework; part~(b) shows a hierarchical drawing computed by our first variant algorithm taking $G$ as input;  part~(c) shows an abstracted hierarchical drawing computed by our final variant that removes all path edges and selected transitive cross edges. Notice that in part~(b) the transitive edges within each vertical path are not shown. %Additionally notice that there are two edges that cross twice. This can be easily rectified at the expense of one extra bend. 
Part~(c) shows a hierarchical drawing where all path edges and transitive cross edges are abstracted.  The advantages of the last drawing are (i) clarity of the drawing due to the sparse representation, (ii) all path edges and transitive edges (within a path) are implied by the $x$ and $y$ coordinates, (iii) the drawn graph has the same transitive closure as $G$, (iv) it gives us a technique to store the transitive closure of $G$ in an extremely compact data structure, and (v) a path between vertices that are on different paths (of the decomposition) can be obtained by traversing one cross edge.

Even though the Sugiyama framework is very popular, and many of the (sub)problems for each phase have turned out to be NP-hard, its main limitation is the fact that the heuristic solutions and decisions that are made during previous phases (e.g., crossing reduction) will influence severely the results obtained in later phases. Nevertheless, previous decisions cannot be changed in order to obtain better results.  This framework can be viewed as a horizontal decomposition of $G$ into (horizontal) layers.  By contrast, the framework of~\cite{ortali2018algorithms}  and all variants presented here can be viewed as a vertical decomposition of $G$ into (vertical) paths/channels. Most problems here are \qmarks{vertically contained} thus reducing their time complexity. It draws either (a) graph $G$ without the transitive \qmarks{path/channel edges} or (b) a condensed form of the transitive closure of $G$. Of course, the \qmarks{missing} incident (transitive) edges of a vertex can be drawn interactively on demand. An added advantage of this framework is that it allows (or it even encourages) the user to use his/her own paths as input to the algorithms.  This means that paths/channels that are important for specific applications can be easily visualized by vertically aligning their nodes.

The algorithms presented in our paper are variants of the path based algorithm presented in~\cite{ortali2018algorithms}. Namely we present seven variants (including the original one) that progressively remove edges, crossings and bends.  Each variant has its own advantages and disadvantages that can exploited in various applications. Furthermore, due to its flexibility, new variants can be created based on the needs of specific applications. We also present experimental results that further demonstrate the power of edge abstraction and their impact on the number of bends, crossings, edge bundling, etc.  Notice that the above variants can be easily modified to work using the concepts of channel decomposition of a DAG and of channel graph as described in~\cite{ortali2018algorithms}.

Our paper is organized as follows: the next section presents necessary knowledge, including a brief description of the basic concepts of the path based algorithm of~\cite{ortali2018algorithms}. In Section 3 we present the variants that are based on the path based algorithm and the metrics of our experiments. Section 4 presents the experimental results and offers a comparison of the pros and cons of each variant with respect to bends, crossings, and clarity. In Section 5 we present our conclusions and interesting open problems.

\section{Overview of the Path Based Framework}
\label{se:overview}

The Path Based Hierarchical Drawing Framework exploits a new approach to visualize directed acyclic graphs that focus on their reachability information~\cite{ortali2018algorithms}.  This framework is orthogonal to the Sugiyama framework in the sense that it is a vertical decomposition of $G$ into (vertical) paths/channels.  Most problems are \qmarks{vertically contained} thus reducing their time complexity. The vertices of a graph $G$ are partitioned into paths, called a \emph{path decomposition} and the vertices of each path are drawn \emph{vertically aligned}. It consists of only two steps: (a) the cycle removal step (if the graph contains directed cycles) and (b) the hierarchical drawing step. 

For the purposes of reachability we propose that Step~(a) follows a simple approach: compute the \emph{Strongly Connected Components (SCC)} of $G$ in linear time and cluster and collapse each SCC into a supernode. Clearly, the resulting graph will be acyclic. This approach has been used in previous papers for various applications, see for example~\cite{DBLP:journals/tse/GansnerKNV93,DBLP:journals/www/LiHZ17,{ortali2018algorithms}}.

Regarding Step (b), the path decomposition may be application defined, user defined or automatically computed by an algorithm.  There are several algorithms that compute a path decomposition of minimum cardinality~\cite{DBLP:journals/siamcomp/HopcroftK73,DBLP:conf/recomb/KuosmanenPGCTM18,DBLP:conf/stoc/Orlin13,DBLP:journals/siamcomp/Schnorr78}. For the rest of this paper, we will assume that the path decomposition is an input to the algorithm along with $G$.  We use an algorithm that computes a path based hierarchical drawing given a DAG $G=(V,E)$ and a path decomposition $S_p$ of~$G$, see~\cite{ortali2018algorithms}. Due to space limitations we describe the algorithm and relevant results in the Appendix.

%\subsection{The Hierarchical Drawing Step}

A \emph{path decomposition} of $G$ is a set of vertex-disjoint paths $S_p= \{P_1,...,P_k\}$ such that every vertex $v\in V$ belongs to exactly one of the paths of $S_p$. A path $P_i\in S_p$ is called a \emph{decomposition path}. The \emph{path decomposition graph} of $G$ associated with a path decomposition $S_p$ is a graph $H=(V,A)$ obtained from $G$ by removing every edge $e=(u,v)$ that connects two vertices on the same decomposition path $P_i\in S_p$ that are not consecutive in the order of $P_i$. An edge of $H$ is a \emph{cross edge} if it is incident to two vertices belonging to two different decomposition paths, else it is a \emph{path edge}. Graph $H$ is obtained from $G$ by removing some transitive edges between vertices in a same path. A \emph{path based hierarchical drawing} of $G$ given $S_p$ is a hierarchical drawing of $H$ where two vertices of $V$ are placed in a same $x$-coordinate if and only if they belong to a same decomposition path $P_i\in S_p$. Algorithm PB-Draw computes a path based hierarchical drawing of $G$. Thus we can read and understand correctly any reachability relation between the vertices of $G$ by visualizing $H$, as shown in Section~\ref{se:thandle} of the Appendix.

%\new{In Figure~\ref{fig:1} we show an example of a drawing computed by Algorithm PB-Draw. In (a) we show the drawing of a graph $G$ as computed by Tom Sawyer Perspectives~\cite{Tom} (a tool of Tom Sawyer Software) which follows the Sugiyama framework. In (b) we show the drawing $\Gamma$ based on $G$ computed by Algorithm PB-Draw.}

Using a path decomposition with a small cardinality may improve the performance of our algorithm in terms of area, bends, number of crossings and computational time. As discussed at the beginning  of this section, computing such a minimum size path decomposition is a well known problem and it provides a great advantage to this framework. Also, the use of the path decomposition concept adds flexibility to the framework, since the paths can be user defined or application specific. The visibility of such important/critical paths is extremely clear in our drawings, since they are all vertically aligned.

\section{Variants, Metrics, and Datasets}
\label{se:variants}
%\subsection{Proposed Framework}
In this section we present the variants of Algorithm~\ref{algo} (see Appendix) that we used for our experiments and the metrics that we considered. We performed two types of experiments: (a) based on %the interesting 
measurements over datasets with respect to the number of bends and crossings \textit{(Variant~0 and Variant~1)} and~(b) based on edge abstraction \textit{(Variant~2, Variant~3, Variant~4, Variant~5, and Variant~6)}.
%In this section we report the results of these two types of experiments.  The experimental results show that both types of variants affect significantly the number of bends and crossings.

All variants use edge bundling as described by Lemma~\ref{lemma:overlap} of Section~\ref{se:thandle} of the Appendix. Refer to Figure~\ref{fig:2}. Namely, all edges that start from vertices of a decomposition path $P$ and go into the same target vertex $v$ bend at the same point. All such edges use the same straight line segment from the bend to vertex $v$. For example, we bundle edges $(21,30)$ and $(28,30)$ by bending them at the same point and by overlapping them from this point to the target vertex, which is vertex $30$. Similarly we do the same for edges $(4,28)$ and $(20,28)$. This type of edge bundling is very useful in the sense that it reduces the total number of bends and crossings, and it reuses some portions of edges.

%Namely, our drawings can be %characterized by:

%\ \\
%\textbf{Edge Bundling:}
%We can shift horizontally by one unit %the position of
%bend $b_e'$ and all the vertices $v$ %and bends $b_e$ such that $X(v) < %X(b_e)$ and  $X(b) > X(b_e)$.
%\newline
%\ \\
%\textbf{Edge Overlapping:} We can define the path decomposition graph differently by removing some transitive cross edges from $H$. For every vertex $v$ we remove the edge \textit{$(u,v)$} if there exists an edge \textit{$(u',v)$} such that \textit{$u'$} and u are in the same decomposition path $P$ and $u$ precedes $u'$ in the order of $P$ (i.e., edges have common source node).

%Similar to that,  for every vertex $v$ we remove the edge \textit{$(u,v)$} if there exists an edge \textit{$(u',v)$} such that \textit{$u'$} and $v$ are in the same decomposition path $P$ and $u'$ precedes $v$ in the order of $P$ (i.e., edges have common target node). 
\subsection{Variants} We present now a suite of drawing techniques, our variants, that are based on Algorithm~\ref{algo}. Our variants can be further customized depending upon the requirements of an application or a user.

\begin{figure}[!ht]
\centering
\subfigure[]{
    \includegraphics[width=0.4\textwidth,height=10.5cm]{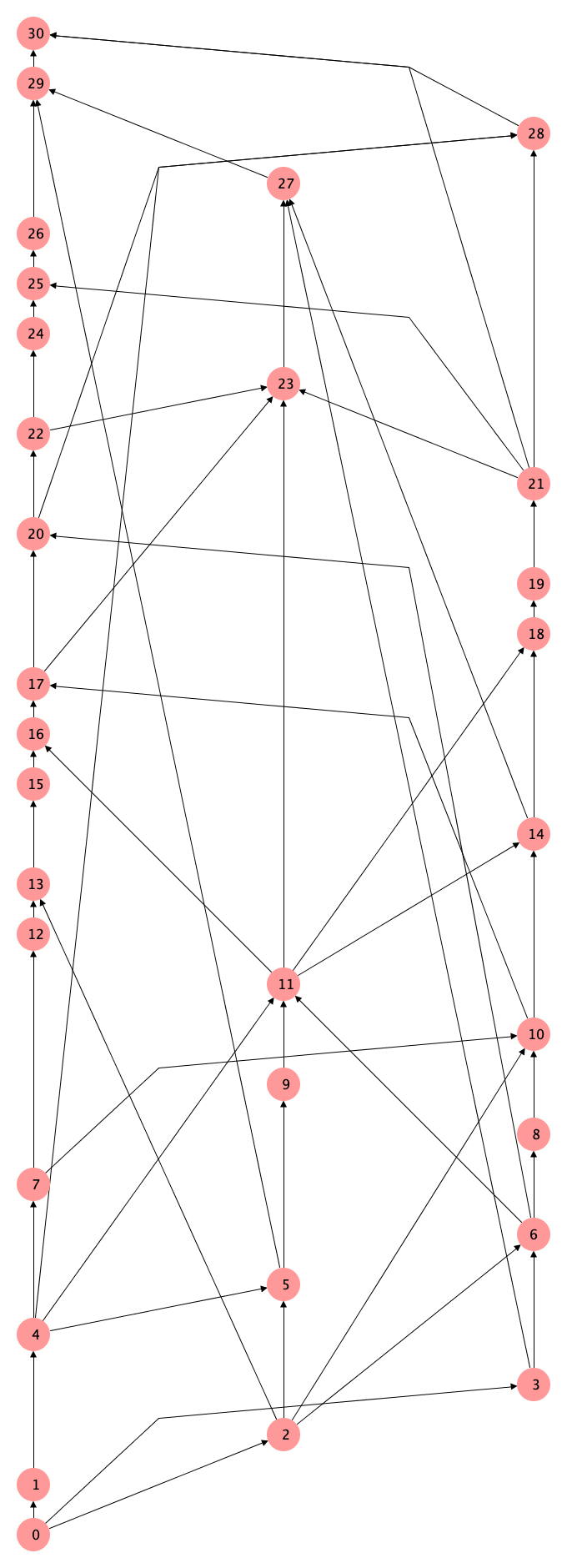}
    \label{fig:2}
}
\subfigure[]{
    \includegraphics[width=0.4\textwidth,height=10.5cm]{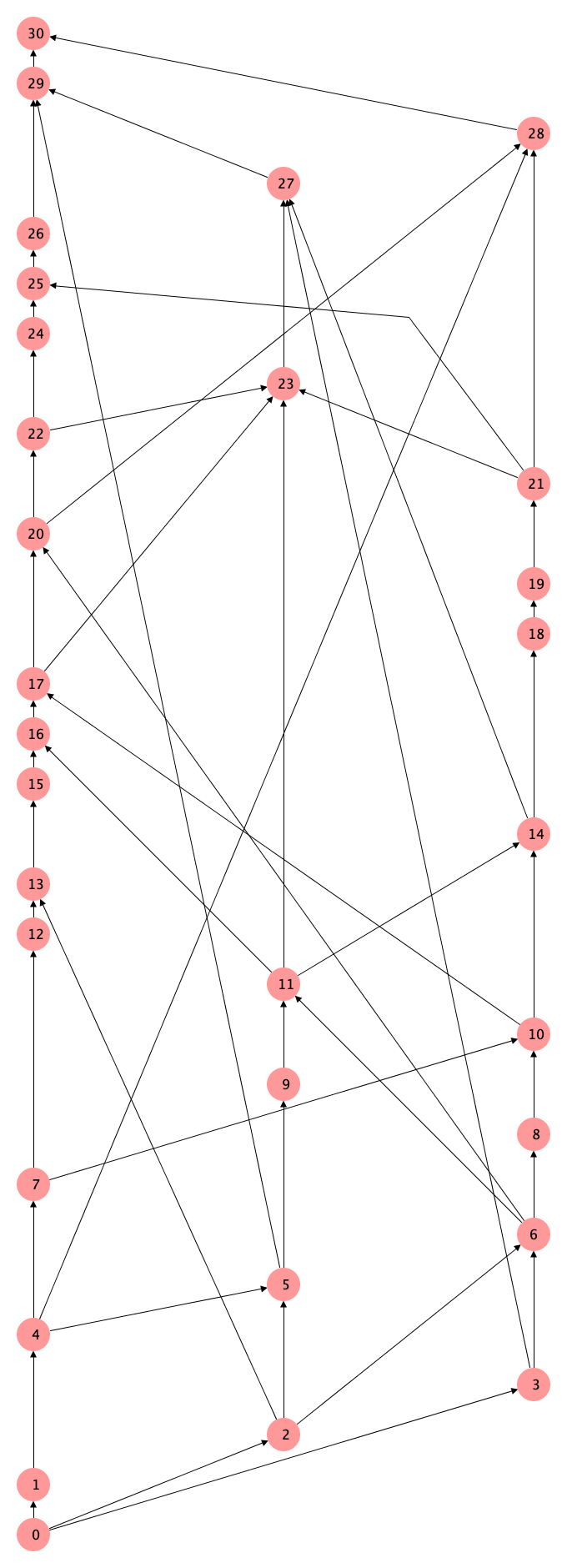}
    \label{fig:3}
}
\caption{ \small{Drawings of DAG~1 drawn with (a)~Variant~1  and (b)~Variant~2.}}
\end{figure}

\begin{figure}[!ht]
\centering
\subfigure[]{
    \includegraphics[width=0.4\textwidth,height=10.5cm]{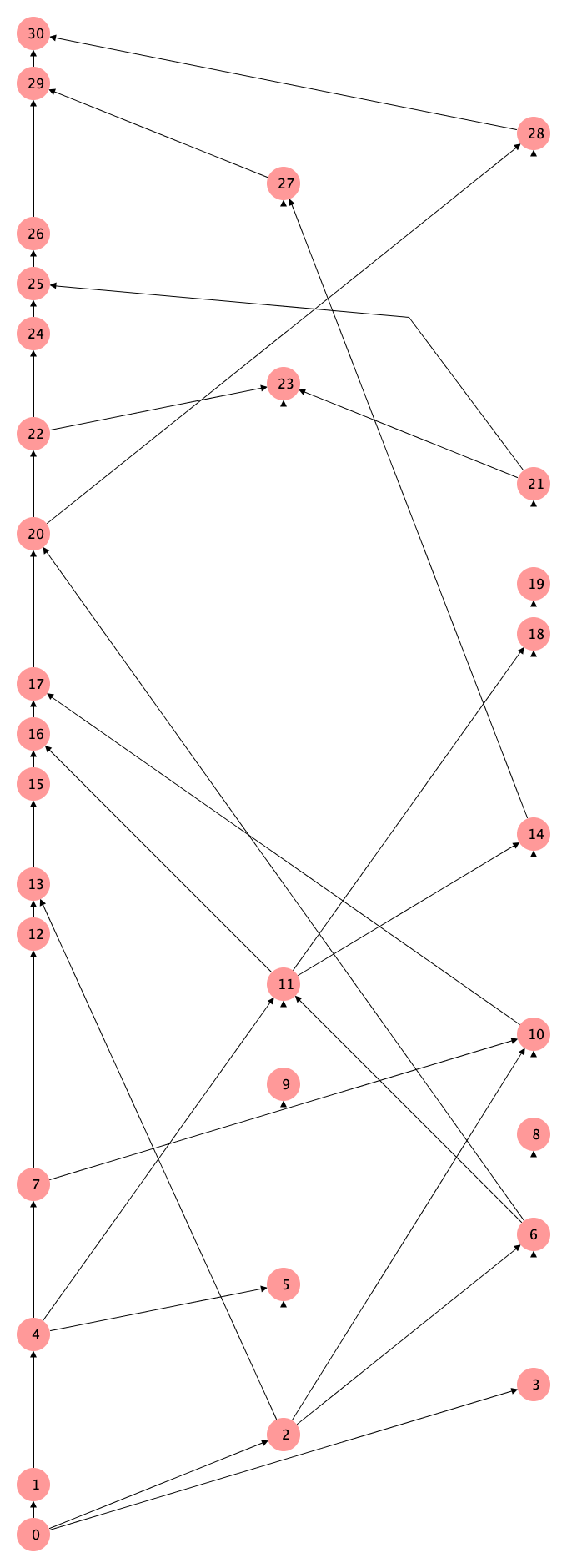}
    \label{fig:4}
}
\subfigure[]{
    \includegraphics[width=0.4\textwidth,height=10.5cm]{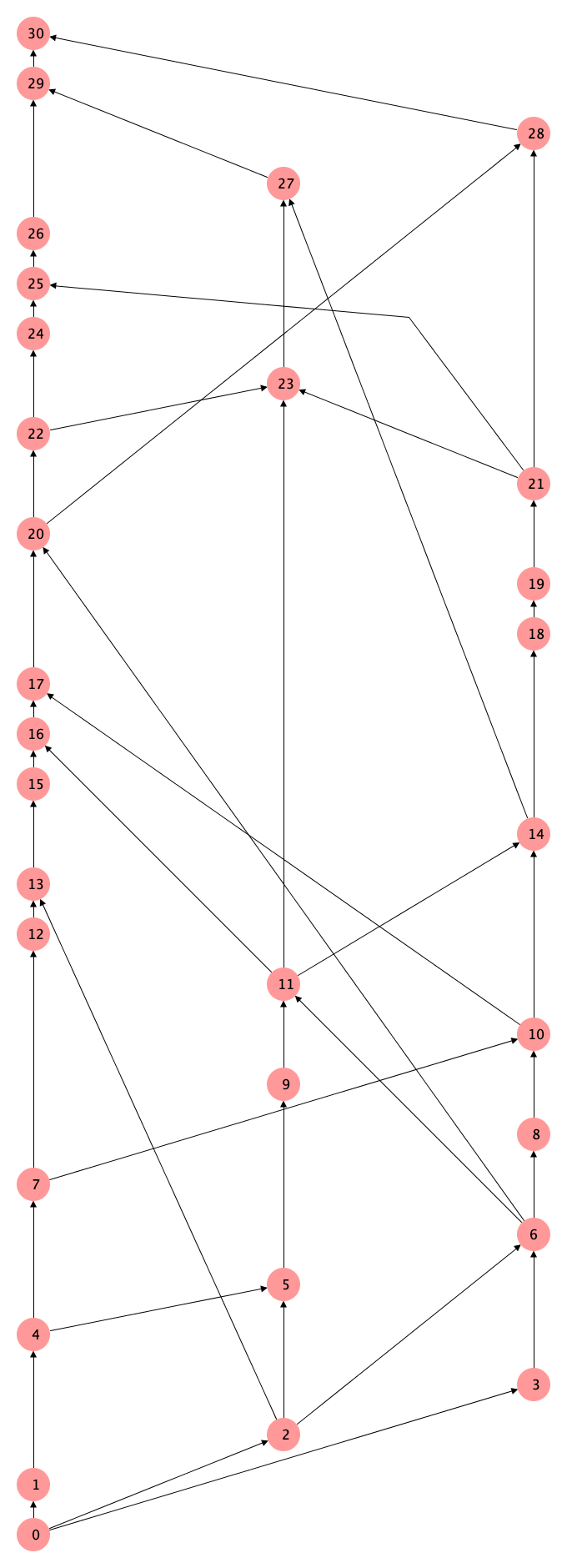}
    \label{fig:5}
}
 \caption{\small{Drawings of DAG~1 drawn with (a)~Variant~3  and (b)~Variant~4.}}
\end{figure}

\begin{itemize}
    \item[$\bullet$] \textbf{Variant 0:} This variant is precisely the same as our baseline, Algorithm~\ref{algo}. See, for example, Figure~\ref{fig:1}.
    \item[$\bullet$] \textbf{Variant 1:} We denote by \emph{jumping cross edge} an edge $e=(u,v)$ such that $|X(v)-X(u)|>1$. In this variant we place a bend on every jumping cross edge of $\Gamma$. Refer Figure~\ref{fig:2}, where, for example, the jumping cross edge $e=(7,10)$ has a bend.
    \item[$\bullet$] \textbf{Variant 2:} For every vertex $u$ we abstract edge \textit{$e_1=(u,v)$} if there exists an edge \textit{$e_2=(u,v')$} such that \textit{$v'$} and $v$ are in the same decomposition path $P$ and $v'$ precedes $v$ in the order of $P$ (edges have common source node). Refer to Figure~\ref{fig:3}, where, for example, $e_1=(2,10)$ and $e_2=(2,6)$. 
    \item[$\bullet$] \textbf{Variant 3:} For every vertex $v$ we abstract the edge \textit{$e_1=(u,v)$} if there exists an edge \textit{$e_2=(u',v)$} such that \textit{$u'$} and $u$ are in the same decomposition path $P$ and $u$ precedes $u'$ in the order of $P$ (edges have common target node). Refer to Figure~\ref{fig:4}, where, for example, $e_1=(21,30)$ and $e_2=(28,30)$.
    \item[$\bullet$] \textbf{Variant 4:} Apply the removal of Variant~2 and Variant~3. Refer to Figure~\ref{fig:5}, where we removed both $(2,10)$ and $(21,30)$.
\end{itemize}

 \subsection{\qmarks{Final Abstraction}}
\label{se:final_ab}
\begin{figure}[!hb]
\centering
\subfigure[]{
    \includegraphics[width=0.4\textwidth,height=10.5cm]{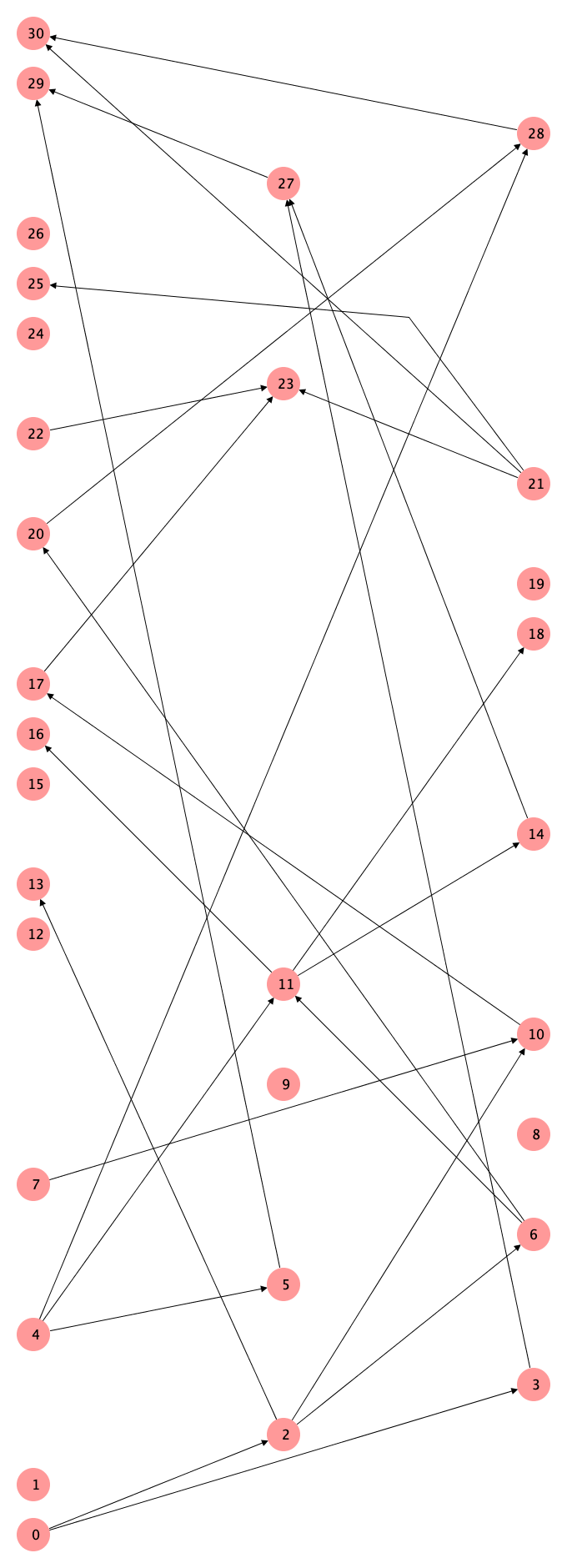}
    \label{fig:6}
}
\subfigure[]{
    \includegraphics[width=0.4\textwidth,height=10.5cm]{vAbstractedd1.png}
    \label{fig:7}
}
\caption{\small{Drawings of DAG~1 drawn with (a)~Variant~5  and (b)~Variant~6.}}
\end{figure}

An important aspect of our work is the preservation of the mental map that can be expressed by the reachability information of a DAG. Since the nodes in each path of the decomposition are vertically aligned, drawing the path edges does not add much information to the mental map of the user.  Hence their removal from the drawing will reduce the number of crossings and the number of edges drawn. Toward to that, we propose an extended abstraction drawing model generated as a combination of the aforementioned variants as shown in Figure~\ref{fig:6} and~\ref{fig:7}.
\par
The main purpose of this abstraction is that we want to retain the visual reachability while minimizing the visual complexity of the drawing. For instance, in our variants as stated in previous sections, paths can be either application based e.g., critical paths or user defined.  Consider Variant 0 the path edges can be removed from the drawing since their existence is implied by the fact that they share the same $x$-coordinate. We refer to this variant as \textit{Variant~5}.
Please notice that we do not remove any number of \qmarks{random} edges in order to create less complex drawings of the same graph but rather we use the unique characteristics of the drawing which may also be application depended. We can further reduce the total number of edges drawn, and as a result the number of crossings, by using this abstraction in combination with \textit{Variant~4} to create a more abstracted drawing, called \textit{Variant~6}. Therefore, we define the following two variants:
\begin{itemize}
    \item[$\bullet$] \textbf{Variant~5}: These drawings are obtained from the drawings of Variant~0 by removing all path edges (see Figure~\ref{fig:6}).
    \item[$\bullet$] \textbf{Variant~6}: These drawings are obtained from the drawings of Variant~4 by removing all path edges (see Figure~\ref{fig:7}).
\end{itemize}

\noindent
\\\\\\
The following theorem is proved in Section~\ref{se:time_variants} of the Appendix.
\begin{theorem}
 \label{th:1,2,3,4,5,6}
     Let $G$ be a DAG with $n$ vertices and $m$ edges, let $S_p$ be a path decomposition of $G$ and let $k$ be the cardinality of $S_p$. It is possible to compute the drawings $\Gamma_1$ according to Variant~1 in $O(n+m)$ time and the drawings $\Gamma_2$, $\Gamma_3$, $\Gamma_4$, $\Gamma_5$, and $\Gamma_6$ according to Variant~2, Variant~3, Variant~4, Variant~5, and Variant~6, respectively, in $O(mk)$ time.
\end{theorem}
\subsection{Metrics and Datasets}
The set of DAGs that was used in the experiments contains five Datasets (DAGs) which were produced in a controlled fashion in order to have a number of nodes and edges, as a factor of the density of the graph. DAG~1 is one of the DAGs that was used to illustrate Algorithm~\ref{algo} in~\cite{ortali2018algorithms}. Table~\ref{datasets} in the Appendix gives a summary for each DAG.

\subsubsection{Metrics for the Experimental Results.} Our analysis aims to evaluate the performance of the various variants of the basic algorithm for each of the aforementioned DAGs. To this end, we use the following:
\begin{description}
\item[$\bullet$ Number of edges drawn in the drawing.]
\item[$\bullet$ Number of cross edges drawn in the drawing.]
\item[$\bullet$ Number of bends.]
\item[$\bullet$ Number of crossings.]
\item[$\bullet$ Execution time:]  is the average execution time for producing each drawing.
\end{description}

\begin{table}[]
\label{table_times}
  \centering
  {\small
  \caption{Average \textbf{execution} times  of the variants  over the \textit{5 DAGs}.\label{executiontime}}
  \begin{tabular}{l|ccccc}
    \hline
    {\small\textit{Variants}}
    & {\small \textit{DAG~1}}
       & {\small \textit{DAG~2}}
      & {\small \textit{DAG~3}}
      & {\small \textit{DAG~4}}
    & {\small \textit{DAG~5}} \\
    \hline
 Variant 0     &36 ms    &51 ms   &59 ms   &103 ms   &137 ms\\
 Variant 1     &33 ms    &40 ms   &55 ms   &114 ms   &128 ms\\
 Variant 2     &37 ms   &39 ms   &62 ms   &99 ms   &138 ms\\
 Variant 3     &41 ms   &48 ms   &46 ms   &93 ms   &141 ms\\
 Variant 4     &49 ms   &43 ms   &54 ms   &108 ms   &104 ms\\
 \hline
  \end{tabular}
  }
\end{table}

\section{Analysis of the Performance}

In this section we analyze the results of the experiments presented in this paper.  We remark that the variants and experiments are described for the path based framework, but they can be used with the channel based framework as well.

Table~\ref{executiontime} shows the performance, \textit{Execution time}~(ms), of the Java implementation  of our suite of drawing solutions as produced by Tom Sawyer Software TS Perspectives~\cite{Tom}. We do not report the execution times of Variant~5 and Variant~6 since they are similar to the execution times of Variant~0 and Variant~4, respectively. We observe that our variants produce hierarchical drawings suitable for large datasets since the reachability information can be seen with little effort while the execution time to produce these results is rather small.

\begin{figure}[!ht]
\includegraphics[width=\textwidth]{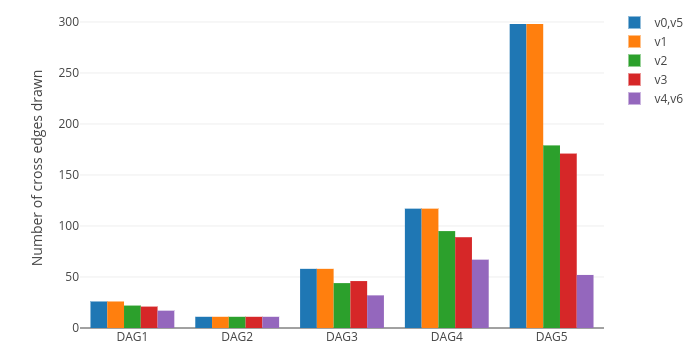}
\caption{Results on \textit{number of cross edges drawn} for each variant over all DAGs.}
\label{m1vd}
\end{figure}

\begin{figure}[!ht]
\includegraphics[width=\textwidth,height=6cm]{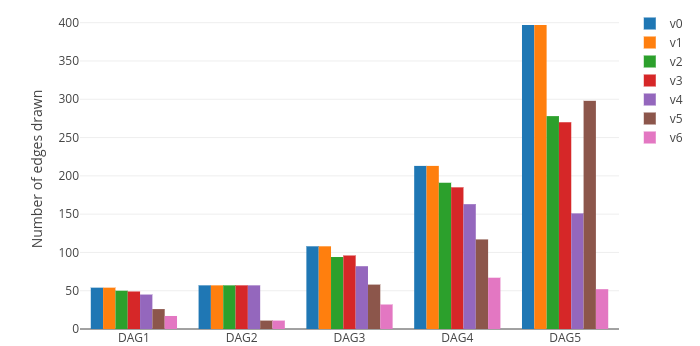}
\caption{Results on \textit{number of edges drawn} for each variant over all DAGs.}
\label{m2vd}
\end{figure}

\begin{figure}[!ht]
\includegraphics[width=\textwidth,height=6cm]{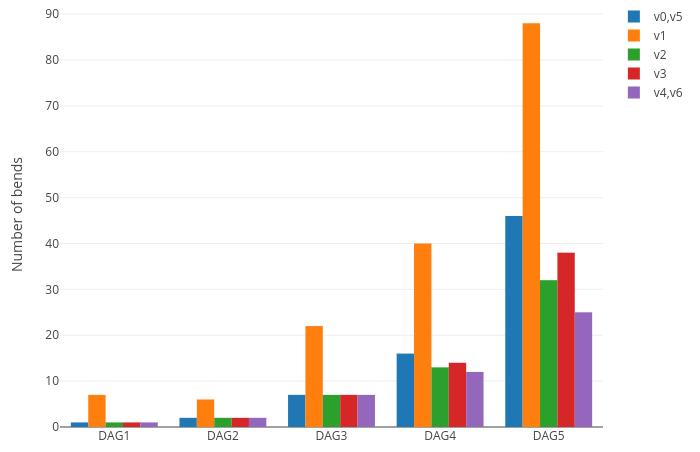}
\caption{Results on \textit{number of bends} for each variant over all DAGs.}
\label{m3vd}
\end{figure}

\begin{figure}[!ht]
\centering
\subfigure[]{
    \includegraphics[width=0.45\textwidth,height=6cm]{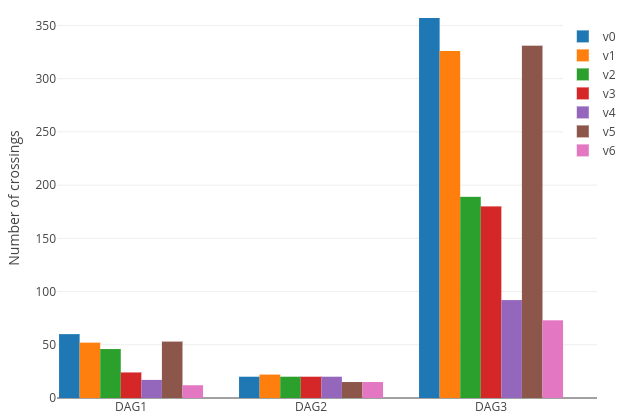}
}
\subfigure[]{
    \includegraphics[width=0.45\textwidth,height=6cm]{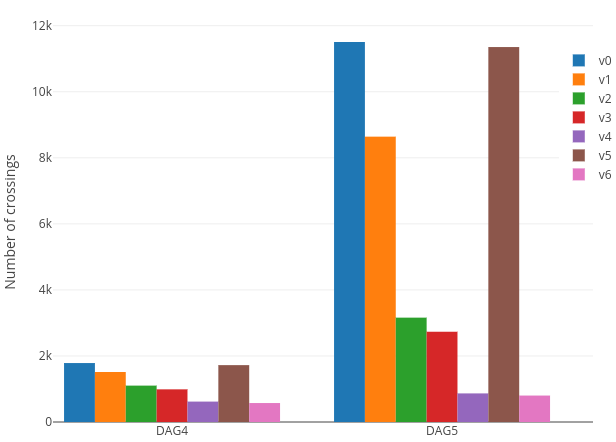}
}
\caption{Results on \textit{number of crossings} for each variant over the (a) DAGs 1,2,3 and (b) DAGs 4,5.}
\label{m4vd}
\end{figure}

The first figure reflects the \textit{the number of edges drawn} for each of the variants over the five DAGs illustrated in Figure~\ref{m1vd}. Similar to that,  Figures~\ref{m2vd}, \ref{m3vd}, and~\ref{m4vd} show the results regarding the \textit{number of cross edges drawn, bends and crossings} respectively for each of the variants. 
\par
 % while its performance in terms of execution time is not significant much higher.

\par
First we discuss the number of edges drawn by our variants. By construction Variant~0 and Variant~1 draw exactly the same set of edges as it is evidenced by Figures~\ref{m1vd} and \ref{m2vd}. The same figures show that Variant~2 and Variant~3 are similar in the number of edges they draw. Clearly, the number of edges drawn by Variant~4 is significantly lower than the number of edges drawn by the other variants. This effect is emphasized in Figure~\ref{m1vd}, where the number of cross edges drawn by Variant~4 for DAG~5 is about one sixth of the number of cross edges drawn by Variant~0 and Variant~1. Finally, we focus on Variant~5 and Variant~6. The sets $E_5$ and $E_6$ of the edges that Variant~5 and Variant~6 draw is a subset of the sets of edges $E_0$ and $E_4$ that Variant~0 and Variant~4 draw, respectively. The cardinality of $E_5$ and $E_6$ is much smaller than the cardinality of $E_0$ and $E_4$ if most of the edges drawn by Variant~0 and Variant~4 are path  edges, as shown in Figure~\ref{m2vd} for DAG~2. Variant~5 and Variant~6 by construction draw the same set of cross edges of respectively Variant~0 and Variant~4.
\par
%The number of bends in drawings computed by Variant~2,  Variant~3, and Variant~4 is rather low. 
As can be seen in Figure~\ref{m3vd} the drawings computed by Variant~2,  Variant~3, and Variant~4 have very few bends on the average. For example, DAG~5 in Variant~3 has 270 edges and the corresponding drawing has only 38 bends, i.e., we have 0.14 bends per edge.
On the other hand, the drawing computed by Variant~1 is less efficient in placing bends. Refer again to DAG~5: in Variant~1 this DAG has 397 edges and the corresponding drawing has 88 bends, i.e, we have 0.22 bends per edge. The number of bends in drawings computed by Variant~5 and  Variant~6 and respectively Variant~0 and  Variant~4 is the same, since the path edges are drawn straight line in all our variants.
\par
The number of crossings is influenced heavily by the number of edges drawn and the extent of edge bundling.  Figure~\ref{m4vd} shows that the performance of Variant~1 is slightly better than that of Variant~0. This can be explained by the fact that in Variant~1 there are more bundles of edges and this naturally decreases the number of crossings. The other variants all have much better performance than Variant~0 and Variant~1 because the corresponding drawings contain significantly fewer edges. Figure~\ref{m4vd} shows that the number of crossings is almost the same in the drawings of Variant~5 and  Variant~6 and Variant~0 and  Variant~4, respectively. This result is very important, since it is an evidence of the fact that path edges participate in a few crossings and, therefore, the decomposition paths can be visualized very clearly in our drawings.

%Finally, Table~\ref{executiontime} shows that the execution time does not vary significantly depending on which variant we choose. 
%
%However, Variant~4 seems to be the most scalable, since the increase of its execution time is modest as the number of edges of the input DAG increases. 
%

\section{Conclusions and Open Problems}
We presented a set of variant algorithms that attempt to draw DAGs hierarchically with few bends and crossings, and by abstracting edges in order to improve the clarity of the drawings. 
\par
Our study assumes that the path decomposition is given as part of the input, or a minimum size decomposition is computed by one of the known algorithms.  However, it is interesting to study the problem of computing a path decomposition and placement of the paths of $G$ which implies the minimum number of jumping cross edges in our drawings. The use of such a decomposition and placement would considerably reduce the number of edges drawn, bends, and crossings in our drawings.  Another open problem is the development and implementation of some compaction strategies, which would improve the readability of our drawings and reduce their height. Finally, it would be important to comprehend human understanding issues related to the removal of some transitive edges and increasing reachability comprehension.


\bibliographystyle{splncs04}
\clearpage
\bibliography{diffBibBnodesConf} 

\begin{thebibliography}{10}
\providecommand{\url}[1]{\texttt{#1}}
\providecommand{\urlprefix}{URL }
\providecommand{\doi}[1]{https://doi.org/#1}

\bibitem{Tom}
Tom {Sawyer} {Software}, \url{www.tomsawyer.com}

\bibitem{DBLP:journals/tse/GansnerKNV93}
Gansner, E.R., Koutsofios, E., North, S.C., Vo, K.: A technique for drawing
  directed graphs. {IEEE} Trans. Software Eng.  \textbf{19}(3),  214--230
  (1993). \doi{10.1109/32.221135}, \url{https://doi.org/10.1109/32.221135}

\bibitem{DBLP:journals/siamcomp/HopcroftK73}
Hopcroft, J.E., Karp, R.M.: An n\({}^{\mbox{5/2}}\) algorithm for maximum
  matchings in bipartite graphs. {SIAM} J. Comput.  \textbf{2}(4),  225--231
  (1973). \doi{10.1137/0202019}, \url{https://doi.org/10.1137/0202019}

\bibitem{Jagadish:1990:CTM:99935.99944}
Jagadish, H.V.: A compression technique to materialize transitive closure. ACM
  Trans. Database Syst.  \textbf{15}(4),  558--598 (Dec 1990).
  \doi{10.1145/99935.99944}, \url{http://doi.acm.org/10.1145/99935.99944}

\bibitem{DBLP:conf/sigmod/JinRDY12}
Jin, R., Ruan, N., Dey, S., Yu, J.X.: {SCARAB:} scaling reachability
  computation on large graphs. In: Proceedings of the {ACM} {SIGMOD}
  International Conference on Management of Data, {SIGMOD} 2012, Scottsdale,
  AZ, USA, May 20-24, 2012. pp. 169--180 (2012). \doi{10.1145/2213836.2213856},
  \url{https://doi.org/10.1145/2213836.2213856}

\bibitem{DBLP:conf/recomb/KuosmanenPGCTM18}
Kuosmanen, A., Paavilainen, T., Gagie, T., Chikhi, R., Tomescu, A.I.,
  M{\"{a}}kinen, V.: Using minimum path cover to boost dynamic programming on
  dags: Co-linear chaining extended. In: Research in Computational Molecular
  Biology - 22nd Annual International Conference, {RECOMB} 2018, Paris, France,
  April 21-24, 2018, Proceedings. pp. 105--121 (2018).
  \doi{10.1007/978-3-319-89929-9\_7},
  \url{https://doi.org/10.1007/978-3-319-89929-9\_7}

\bibitem{DBLP:journals/www/LiHZ17}
Li, L., Hua, W., Zhou, X.: {HD-GDD:} high dimensional graph dominance drawing
  approach for reachability query. World Wide Web  \textbf{20}(4),  677--696
  (2017). \doi{10.1007/s11280-016-0407-z},
  \url{https://doi.org/10.1007/s11280-016-0407-z}

\bibitem{DBLP:conf/stoc/Orlin13}
Orlin, J.B.: Max flows in {O}(nm) time, or better. In: Symposium on Theory of
  Computing Conference, STOC'13, Palo Alto, CA, USA, June 1-4, 2013. pp.
  765--774 (2013). \doi{10.1145/2488608.2488705},
  \url{http://doi.acm.org/10.1145/2488608.2488705}

\bibitem{ortali2018algorithms}
Ortali, G., Tollis, I.G.: Algorithms and bounds for drawing directed graphs.
  In: International Symposium on Graph Drawing and Network Visualization. pp.
  579--592. Springer (2018)

\bibitem{DBLP:conf/sigmod/SchaikM11}
van Schaik, S.J., de~Moor, O.: A memory efficient reachability data structure
  through bit vector compression. In: Proceedings of the {ACM} {SIGMOD}
  International Conference on Management of Data, {SIGMOD} 2011, Athens,
  Greece, June 12-16, 2011. pp. 913--924 (2011). \doi{10.1145/1989323.1989419},
  \url{https://doi.org/10.1145/1989323.1989419}

\bibitem{DBLP:journals/siamcomp/Schnorr78}
Schnorr, C.: An algorithm for transitive closure with linear expected time.
  {SIAM} J. Comput.  \textbf{7}(2),  127--133 (1978). \doi{10.1137/0207011},
  \url{https://doi.org/10.1137/0207011}

\bibitem{sugiyama1981methods}
Sugiyama, K., Tagawa, S., Toda, M.: Methods for visual understanding of
  hierarchical system structures. IEEE Transactions on Systems, Man, and
  Cybernetics  \textbf{11}(2),  109--125 (1981)

\bibitem{DBLP:conf/edbt/VelosoCJZ14}
Veloso, R.R., Cerf, L., Jr., W.M., Zaki, M.J.: Reachability queries in very
  large graphs: {A} fast refined online search approach. In: Proceedings of the
  17th International Conference on Extending Database Technology, {EDBT} 2014,
  Athens, Greece, March 24-28, 2014. pp. 511--522 (2014).
  \doi{10.5441/002/edbt.2014.46},
  \url{https://doi.org/10.5441/002/edbt.2014.46}

\bibitem{DBLP:journals/vldb/YildirimCZ12}
Yildirim, H., Chaoji, V., Zaki, M.J.: {GRAIL:} a scalable index for
  reachability queries in very large graphs. {VLDB} J.  \textbf{21}(4),
  509--534 (2012). \doi{10.1007/s00778-011-0256-4},
  \url{https://doi.org/10.1007/s00778-011-0256-4}

\end{thebibliography}
\chapter*{Appendix}

\section{Algorithm PB-Draw}
\label{se:thandle}

The following algorithm and theorems and lemmas are from~\cite{ortali2018algorithms}

\begin{theorem}
	\label{theorem:HH'}
	Let $G$ be a DAG and let $S_p$ be a path decomposition of $G$. The path based graph $H$ of $G$ associated with $S_p$ have the same reachability properties of the $G$.
\end{theorem}
\noindent

\begin{algorithm}
\caption{PB-Draw($S_p$, $H$)\newline
{\bf Input:}   a path decomposition $S_p$ of a DAG $G$; a path based graph $H$ of $G$ associated with $S_p$; a topological sorting $T$ of the vertices of $H$, where $T_v$ is the position of $v$ in this sorting.
\newline
{\bf Output:}  The path based hierarchical drawing $\Gamma$ of $G$ associated with $S_p$.}
\label{algo}
\begin{algorithmic}[1]
    \item Compute a drawing $\Gamma$ of $H$ by: 
    \begin{enumerate}
        \item assign to every vertex $v\in V$ belonging to the decomposition path $P_i$ an $x$-coordinate $X(v)=2i$ and an $y$-coordinate $Y(v)=T_v$
        \item  draw every edge $e=(u,v)\in E$ straight line. 
    \end{enumerate}
    \item If the straight line drawing of $e$ intersects some vertex $w\in V$ different from $u$ or $v$ in $\Gamma$, introduce a bend on $e$ in the position $(X_b,Y_b)$, where $Y_b=Y(v)-1$ and:
    \begin{enumerate}
    \item  If $X(u)<X(v)$: $X_b=X(u)+1$
    \item Else, if $X(u)\ge X(v)$: $X(b_e)=X(u)-1$
    \end{enumerate}
 \end{algorithmic}
\end{algorithm}

Let $\Gamma$ be a drawing computed by PB-Draw. The following Lemmas and Theorem are proved in~\cite{ortali2018algorithms}.

\begin{lemma}
	\label{lemma:no_intersection}
	Any edge $e=(u,v)$ does not intersect a vertex different from $u$ and $v$ in $\Gamma$.
\end{lemma}
The following lemma shows how Algorithm PB-Draw bundles the edges. We apply this technique of bundling in every variant of PB-Draw that we introduce in the next sections; for this reason, we better describe it in Section~\ref{se:variants}.
\begin{lemma}
	\label{lemma:overlap}
	Let $e=(u,v)$ and $e'=(u',v')$ be two edges drawn with a bend in~$\Gamma$. Their bends are placed in the same point if and only if $u$ and $u'$ are in the same decomposition path and $v=v'$.
\end{lemma}

Finally, the following theorem shows that Algorithm PB-Draw computes efficiently path hierarchical drawings with a very small area, number of bends and number of bends per edge.
\begin{theorem}
	\label{theorem:area}
	Let $G$ be a DAG with $n$ vertices and $m$ edges, let $S_p$ be a path decomposition of $G$ and let $k$ be the cardinality of $S_p$. Algorithm PB-Draw computes a path based hierarchical drawing $\Gamma$ of $G$ given $S_p$ in $O(mk)$ time. Furthermore, $Area(\Gamma)=O(kn)$ and every edge has at most one bend.
\end{theorem}

\section{ Proof of Theorem \ref{th:1,2,3,4,5,6}}
\label{se:time_variants}

In this section we prove Theorem \ref{th:1,2,3,4,5,6}. First of all, we study the case of the drawing $\Gamma_1$ computed according to Variant~1. We can compute $\Gamma_1$ by simply ignoring Step~2 of Algorithm~\ref{algo}, which requires $O(mk)$ time, and placing a bend in every jumping cross edge. This operation and all the other operations of Algorithm~\ref{algo} require $O(n+m)$ time. Hence, we can compute $\Gamma_1$ in $O(n+m)$ time. We now focus our attention on Variants~2-6.

Let $G$ be a DAG and let $S_p$ be a path decomposition of $G$ having cardinality $k$. Let $\Gamma_0$ be a drawing of $G$ computed by Algorithm~\ref{algo}. We present Algorithm Compute-V2, which describes how it is possible to compute a drawing $\Gamma_2$ according to Variant~2 given $\Gamma_0$. The algorithm  takes as input a path decomposition $S_p$ of $G$ and $\Gamma_0$ and it gives as output $\Gamma_2$. For every vertex of $\Gamma_0$ the algorithm checks the adjacent vertices and, for every path $P_i$, it stores the one having minimum $y$-coordinate  in an array $A_k$. Then it removes all edges $(v,w)$ such that $Y(w)$ is not stored in $A_k$.
 \begin{algorithm}
\caption{Compute-V2 ($S_p$,$\Gamma_0$)\newline
{\bf Input:}   a path decomposition $S_p=(P_1,...,P_k)$ of a DAG $G$; a path based hierarchical drawing $\Gamma_0$ according to Variant~0
\newline
{\bf Output:}  a path based hierarchical drawing $\Gamma_2$ according to Variant~2}
\label{algoV2}
\begin{algorithmic}[1]
    \item $\Gamma_2= \Gamma_0$
    \item For any vertex $v$ drawn in $\Gamma_0$: 
    \begin{itemize}
        \item Create an array $A$ of $k$ positions
        \item For every edge $e=(v,w)$ in $\Gamma_0$ 
        \begin{itemize}
        \item Let $P_i$ be the path of $w$
        \item If $A[i]=void$:
        \begin{itemize}
            \item $A[i]=Y(w)$
        \end{itemize}
        \item Else:
         \begin{itemize}
            \item $A[i]=min(A[i],Y(w))$
        \end{itemize}
        
    \end{itemize}
    \end{itemize}
    \begin{itemize}
        \item For every edge $e=(v,w)$ 
        \begin{itemize}
        \item Let $P_i$ be the path of $w$
        \item If $T_w \neq A_i$
        \begin{itemize}
            \item remove $e$ from $\Gamma_2$
        \end{itemize}
    \end{itemize}
    \end{itemize}
 \end{algorithmic}

\end{algorithm}
 
 Algorithm Compute-V2 requires linear time, since it visits every vertex once and every edge twice and, for every visit, it performs a constant number of operations.
 
 It is easy to see that we can define two similar algorithms in order to compute drawings $\Gamma_3$ and $\Gamma_4$ according to  Variant~3 and Variant~4, respectively, by taking $\Gamma_0$ and $S_p$ as input. Therefore, given $\Gamma_0$ it is possible to compute $\Gamma_2$, $\Gamma_3$ and $\Gamma_4$ in $O(n+m)$ time. It is also true for the drawings $\Gamma_5$ and $\Gamma_6$ according to Variant~5 and Variant~6, since they can be computed from $\Gamma_0$ and $\Gamma_4$ by removing the path edges. Moreover, notice that $\Gamma_0$ can be computed in $O(km)$, according to Theorem~\ref{theorem:area}. Hence, we can compute $\Gamma_2$, $\Gamma_3$, $\Gamma_4$, $\Gamma_5$, $\Gamma_6$ in $O(mk)$.  \qed

%The following theorem  is a direct consequence of Theorem~\ref{theorem:area}, Theorem~\ref{th:1,2,3,4}, and of the definitions of Variant~5 and Variant~6.

 %\begin{theorem}
 %\label{th:5,6}
   %  Let $G$ be a DAG with $n$ vertices and $m$ edges, let $S_p$ be a path decomposition of $G$ and let $k$ be the cardinality of $S_p$. It is possible to compute drawings $\Gamma_5$ and $\Gamma_6$ according to  Variant~5 and Variant~6, respectively, in $O(kn+m)$ time.
%\end{theorem}

\section{Experiment Details}
Here we report the experimental details. The experiments  run on a single machine having an 3.1 Ghz i7 dual core, 16 GB main memory and 500GB flash storage disk space. We report the average time of 5 runs per DAG.

\begin{table}[]
\label{table_dataset}
\centering
\caption{DAGs Statistics.}\label{datasets}
\begin{tabular}{|l|l|c|}
\hline
Name of Dataset &  Number of Nodes and Edges & Completeness (\%)\\
\hline
DAG~1 & {\itshape 30 nodes and 69 edges}  & \	$\sim$ 16\\
DAG~2 & {\itshape 50 nodes and 61 edges} & 5\\
DAG~3 & {\itshape 50 nodes and 121 edges} & 10\\
DAG~4 & {\itshape 100 nodes and 246 edges} & 5\\
DAG~5 & {\itshape 100 nodes and 494 edges} & 10\\
\hline
\end{tabular}
\end{table}
\end{document}